\documentclass[12pt]{article}
\usepackage{amssymb,amsmath,graphicx}
\begin{document}
\title{\textbf{Mass gap without vacuum energy}} 
\author{B.~Holdom%
\thanks{bob.holdom@utoronto.ca}\\
\emph{\small Department of Physics, University of Toronto}\\[-1ex]
\emph{\small Toronto ON Canada M5S1A7}}
\date{}
\maketitle
\begin{abstract}
We consider soft nonlocal deformations of massless theories that introduce a mass gap. By use of a renormalization scheme that preserves the ultraviolet softness of the deformation, renormalized quantities of low mass dimension, such as normal mass terms, vanish via finite counterterms. The same applies to the renormalized cosmological constant. We connect this discussion to gauge theories, since they are also subject to a soft nonlocal deformation due to the effects of Gribov copies. These effects are softer than usually portrayed.
\end{abstract}

We wish to explore the apparently tight relationship between the existence of a mass gap and the existence of vacuum energy. Consider first the case of a free massless scalar field. By scale invariance $T_\mu^\mu=0$ and Lorentz invariance $\langle T_{\mu\nu}\rangle\propto \eta_{\mu\nu}$ we know that the vacuum energy vanishes. There is an apparent contribution from the zero-point energies that requires regularization, but a regulator in the form of an ultraviolet cutoff breaks both the scale and Lorentz invariance. A regulator that preserves scale invariance, such as dimensional regularization, automatically produces the expected vanishing of $\langle T_{\mu\nu}\rangle$. This continues to be true in perturbation theory in an interacting $\lambda\phi^4$ theory, as long as there is no dimensionful coupling in the theory.

The situation changes with the introduction of an explicit mass term $m^2\phi^2$. Now an infinite cosmological constant counterterm proportional to $m^4$ is required in the renormalization procedure (see \cite{n1} for a thorough discussion). This occurs even in the free theory where the result for the renormalized cosmological constant is
\begin{equation}
\Lambda_r=\frac{m^4}{32\pi^2}\log(m/\mu)
.\end{equation}
The renormalization scale $\mu$ also absorbs a renormalization scheme dependence. At any order in perturbation theory in the interacting theory there is a similar result in terms of renormalized quantities,
\begin{equation}
\Lambda_r=m^4 f_\Lambda(\lambda,m/\mu)
.\end{equation}
Vacuum energy and mass appear to be inextricably linked.

Our focus must therefore remain on interacting theories that have no explicit mass parameters. The scale anomaly in such theories involves only dimension four operators $T^\mu_\mu=\sum_i\beta_i(\lambda){\cal O}_i$. Suppose such a theory develops a mass gap through some dynamical or nonperturbative means.  It would usually be expected that the mass gap $M$ would imply that $\langle T^\mu_\mu\rangle\approx M^4$ which in turn implies that at least some combination of dimension four operators develop vacuum expectation values. The operator product expansion translates these vevs into power law corrections to the ultraviolet behavior of the theory, with these corrections of the form $M^4/p^4$. Thus if we want to contemplate a mass gap that could develop without vacuum energy, then the power law corrections that result in the ultraviolet will have to be softer than this.

To explore the consistency of this possibility we extend the scalar theory through the addition of a few nonlocal terms so that the bare Lagrangian in Euclidean form is
\begin{equation}
{\cal L}_E=\frac{1}{2}(\partial_\mu\phi)^2+\frac{\lambda}{4!}\phi^4+\frac{1}{2}m^2\phi^2+\Lambda+\frac{1}{2}\overline{\Lambda}\frac{1}{\square}\phi^2+\frac{1}{2}\kappa^6\frac{1}{\square}\frac{1}{\square}\phi^2
.\end{equation}
The nonlocal terms have the effect of damping $\phi$ fluctuations in the infrared, while producing small, subleading power law corrections in the ultraviolet. The $\overline{\Lambda}$ notation emphasizes that this parameter has the same dimension as $\Lambda$. $\overline{\Lambda}\neq0$ produces a $\overline{\Lambda}/p^4$ power law correction in the propagator and an infinite contribution to $\Lambda$ very similar to the $m^2\neq0$ case. More interesting is the softer ultraviolet power law correction $\kappa^6/p^6$ from the last term in the case $m^2=\overline{\Lambda}=0$. Then the Euclidean propagator is
\begin{equation}
\Delta(p)=\frac{1}{p^2+\frac{\kappa^6}{p^4}}
,\label{e1}\end{equation}
which has a massive pole in Minkowski space. We refer to this mass gap as ultrasoft, and in this case the resulting contribution to $\Lambda$ is at most finite.

We need a renormalization scheme that ensures that dimensionful couplings of mass dimension less than six are not generated. In dimensional regularization, the infinite counterterms that arise are proportional to integer powers of the couplings in the theory. If $\kappa^6$ is the only dimensionful coupling then the counterterms as represented by the bare parameters $m^2$, $\overline{\Lambda}$ and $\Lambda$ remain finite. We require that the corresponding renormalized quantities vanish order by order in perturbation theory. At zeroth order,
\begin{equation}
\Lambda_r^{(0)}=\frac{\pi}{2\sqrt{3}}\frac{\kappa^4}{(4\pi)^2}+\Lambda^{(0)}=0
\end{equation}
while at first order,
\begin{equation}
(m_r^2)^{(1)}=-\frac{\pi}{3\sqrt{3}}\frac{\lambda\kappa^2}{(4\pi)^2}+(m^2)^{(1)}=0
,\end{equation}
\begin{equation}
\Lambda_r^{(1)}=\frac{\pi^2}{54}\frac{\lambda\kappa^4}{(4\pi)^4}+\Lambda^{(1)}=0
.\end{equation}
The renormalization of $\overline{\Lambda}$ can first occur at order $\lambda^2$,
\begin{equation}
\overline{\Lambda}_r^{(2)}={\cal O}(\lambda^2\kappa^4)+\overline{\Lambda}^{(2)}=0
.\end{equation}
The ${\cal O}(\lambda^2\kappa^4)$ quantity would be obtained by isolating a $\kappa^4/p^4$ power law correction to the large $p^2$ behavior of the 2-point function. Clearly this renormalization procedure can be continued to yield $m_r^2=\overline{\Lambda}_r=\Lambda_r=0$ at any order.

This defines an ultrasoft theory. It has vanishing vacuum energy and a propagator with ultraviolet behavior
\begin{equation}
\lim_{p^2\rightarrow\infty}[p^2\Delta_r(p^2)]^{-1}= f_\Delta(\lambda,p/\mu)(1+{\cal O}(\frac{\kappa^6}{p^6}))
.\end{equation}
The only infinite renormalization of the 2-point function is the standard wave function renormalization as reflected by the $\mu$ dependence of $f_\Delta$. Other than the existence of a mass gap our considerations are not constraining the form of the full propagator at small $p^2$, which could differ substantially from the form of the zeroth order propagator in (\ref{e1}).

We now present another example of a very soft deformation which may be of more immediate interest for gauge theories. First we put the massless interacting scalar field theory in a finite volume $V$. The point will be to constrain the discrete set of fourier amplitudes $\phi(p)$,
\begin{equation}
\phi(p)=\frac{1}{\sqrt{V}}\int_V d^4x \phi(x) e^{-i p x}
,\end{equation}
to lie within a hypercube. This corresponds to adding a highly nonlocal potential term to the theory.
\begin{equation}
V_\kappa(\phi)=\left\{\begin{array}{l}0\quad\quad\quad\mbox{if}\quad({\rm |Re}\phi(p)|<1/\kappa\mbox{ and } |{\rm Im}\phi(p)|<1/\kappa)\quad\forall\quad p\\\infty\quad\quad\mbox{ otherwise}\end{array}\right.
\label{e2}\end{equation}
Alternatively this can be implemented as a constraint on the path integral definition of the generating functional
\begin{equation}
Z_\kappa[J]=e^{-\int_V d^4x \frac{\lambda}{4!}\left(\frac{\delta}{\delta J(x)}\right)^4}\int_{\kappa}{\cal D}\phi(p)\;e^{-\sum_p (p^2\phi(p)^*\phi(p)+J(p)^*\phi(p))}
,\end{equation}
where
\begin{equation}
\int_{\kappa}{\cal D}\phi(p)=\prod_p \left[\int_{-\frac{1}{\kappa}}^\frac{1}{\kappa}d {\rm Re}\phi(p)\int_{-\frac{1}{\kappa}}^\frac{1}{\kappa}d {\rm Im}\phi(p)\right]
,\end{equation}
and
\begin{equation}
J(p)=\frac{1}{\sqrt{V}}\int_V d^4x J(x)e^{-ipx}
.\end{equation}

The propagator becomes
\begin{equation}
\Delta_\kappa(x-y)=\frac{1}{V}\sum_p e^{ip(x-y)}\Delta_\kappa(p)
,\end{equation}
where
\begin{equation}
\Delta_\kappa(p)=\left.\frac{1}{Z_\kappa(p,j)}\frac{d^2}{d j^2}Z_\kappa(p,j)\right|_{j=0}
,\end{equation}
and
\begin{equation}
Z_\kappa(p,j)=\int_{-\frac{1}{\kappa}}^\frac{1}{\kappa}dx \exp(-\frac{1}{2}p^2x^2+jx)
.\end{equation}
We find
\begin{equation}
\Delta^{\rm A}_\kappa(p)=
\left\{\begin{array}{l}\displaystyle\frac{1}{3\kappa^2}-\frac{2}{45}\frac{p^2}{\kappa^4}+...\quad\quad\quad\quad\quad\quad\quad\quad\quad\;\; p^2\ll\kappa^2\\\vspace{-1ex}\\\displaystyle\frac{1}{p^2}\left(1-\sqrt{\frac{2}{\pi}}\frac{\sqrt{p^2}}{\kappa}\exp(-\frac{1}{2}\frac{p^2}{\kappa^2})+...\right)\quad p^2\gg\kappa^2\end{array}\right.
\end{equation}
This propagator (labelled to distinguish it from others below) acts like a massive propagator in the infrared, but it approaches a massless propagator in the ultraviolet exponentially quickly.

Thus restricting the field space to a hypercube has caused an infrared deformation of the massless theory, sufficient to produce a mass gap. It is an extreme version of the previous ultrasoft example, where now the approach to massless behavior in the ultraviolet is faster than any negative power of $p^2$. For this ``infinite softness'' to survive in perturbation theory, finite adjustments of an infinite number of nonlocal terms quadratic in the fields would be required. This again defines a renormalization scheme.

This seemingly artificial example may be relevant to gauge theories. The existence of Gribov copies dictates a similar constraint on the functional integral of gauge theories, to avoid the multiple counting of gauge equivalent configurations \cite{n2}. The gauge field configuration space must be restricted to the fundamental modular region (FMR) \cite{n3}, where this is a bounded convex region within the gauge-fixed hypersurface $\partial_\mu A^\mu=0$. Each gauge in-equivalent configuration occurs once and only once in this region. The boundary of this region is nontrivial and difficult to work with, but the FMR is known to lie within and share part of its boundary with the first Gribov region defined by positive Fadeev-Popov operator ${\cal D}_A\geq0$. This region has an ellipsoidal shape and if translated to our scalar field example would take the form $\sum_p|\phi(p)|^2/p^2\leq C$. A hyperbox that most resembles this region is given by $({\rm |Re}\phi(p)|<p/\kappa^2\mbox{ and } |{\rm Im}\phi(p)|<p/\kappa^2)$ for some $\kappa$. Using this hyperbox rather than the hypercube in (\ref{e2}) yields the following propagator.
\begin{equation}
\Delta^{\rm B}_\kappa(p)=
\left\{\begin{array}{l}\displaystyle\frac{p^2}{3\kappa^4}-\frac{2}{45}\frac{p^6}{\kappa^8}+...\quad\quad\quad\quad\quad\quad\quad\quad\;\;\, p^2\ll\kappa^2\\\vspace{-1ex}\\\displaystyle\frac{1}{p^2}\left(1-\sqrt{\frac{2}{\pi}}\frac{p^2}{\kappa^2}\exp(-\frac{1}{2}\frac{p^4}{\kappa^4})+...\right)\quad p^2\gg\kappa^2\end{array}\right.
\end{equation}
When compared to $\Delta^{\rm A}_\kappa(p)$ this propagator is suppressed even more in the infrared, and approaches $1/p^2$ even more quickly in the ultraviolet.

Since the hyperbox in field space is suppressing the field fluctuations we have the relation $\Delta^{\rm B}_\kappa(p)<1/p^2$. We may use these results to constrain the behavior of the propagator in the case when the constraint in configuration space is not completely known or too difficult to specify, as is the case for the FMR of a gauge theory. In particular there is a smallest possible value of $\kappa$ that would still have the hyperbox completely contained within the FMR. Upon moving the boundary outward from the hyperbox to the convex FMR boundary, the field constraint is being relaxed and the fluctuations are less suppressed. Thus the corresponding propagator $\Delta_{\rm FMR}(p)$ (the scalar factor defining the gauge field propagator) can only be closer to $1/p^2$ and we have
\begin{equation}
\Delta^{\rm B}_\kappa(p)<\Delta_{\rm FMR}(p)<1/p^2
.\end{equation}
This is a very tight constraint on $\Delta_{\rm FMR}(p)$ in the ultraviolet.

There is a complementary approach to the problem of Gribov copies. Rather than attempting to constrain the gauge field configuration space to the FMR, one can consider an equivalent definition of the generating functional of a gauge theory,
\begin{equation}
{\cal Z}[J]=\int [dA] e^{iS[A,J]}\delta(\partial_\mu A^\mu)\left|\det({\cal D}_A)\right|\frac{1}{1+N(A)}
.\end{equation}
As first introduced by Gribov, no restriction to the FMR is needed here since the counting factor $N(A)$ accounts for the number of gauge equivalent copies of a configuration. By explicitly counting Gribov copies \cite{n4} for a class of configurations it is found that the last factor strongly suppresses field configurations outside a certain region in configuration space. Within this region, which in turn lies within the FMR, $N(A)$ identically vanishes. The corresponding bound on the $p$th fourier amplitude is found to be proportional to $p$, which agrees with the shape of the ellipsoid or hyperbox described above. Outside this region $N(A)$ is roughly proportional to the fourier amplitude divided by $p$ all raised to some power $a>3$. Thus even though the $1/(1+N(A))$ factor does not produce a sharp cutoff in field space, it does have a similar effect on the propagator. The result from \cite{n4} is
\begin{equation}
\Delta^{\rm C}_\kappa(p)=
\left\{\begin{array}{l}\displaystyle\frac{1}{3}\frac{a-1}{a-3}\frac{p^2}{\kappa^4}+...\quad\quad\quad\quad\quad\quad\quad\quad\quad\quad\, p^2\ll\kappa^2\\\vspace{-1ex}\\\displaystyle\frac{1}{p^2}\left(1-a\sqrt{\frac{2}{\pi}}\frac{\kappa^2}{p^2}\exp(-\frac{1}{2}\frac{p^4}{\kappa^4})+...\right)\quad p^2\gg\kappa^2\end{array}\right.
\end{equation}

These considerations are showing that the nonlocal effects of Gribov copies in gauge theories are severe in the infrared while being infinitely soft in the ultraviolet. But this ultraviolet softness is not apparent in the well known Gribov-Zwanziger (GZ) approach \cite{n1,n5}. In that semi-perturbative approach the constraint on the gauge fields is implemented through a modification of the action and the result is the appearance of a new nonlocal term, which happens to be of the form of our $\overline{\Lambda}$ term. The resulting propagator is similar to $\Delta^{\rm B}_\kappa(p)$ or $\Delta^{\rm C}_\kappa(p)$ in the infrared, but it differs in the ultraviolet where it instead has a power law correction. We note that the modified action in the GZ approach is arrived at by focusing on the first Gribov region rather than the FMR, and then by requiring that the functional integral is self-consistently dominated by field configurations at the boundary of this region. But some correlation functions are not dominated by such configurations, most notably those evaluated at high $p^2$. The relevant fluctuations are perturbative and their variance is $\sim1/p^2$, and  when $p^2\gg\kappa^2$ these fluctuations are much smaller and far away from the boundaries located at $\sim p^2/\kappa^4$. Since these fluctuations have close to a Gaussian distribution the effects of the boundaries on these fluctuations are expected to be exponentially suppressed, in agreement with what we have seen and in contrast to the GZ prescription.

The softness of the effects of Gribov copies raises a question. Could a pure gauge theory, confining and asymptotically free, be an example of an infinitely soft theory with a mass gap? Apparently not, since vacuum energy receives a nonperturbative contribution in a pure gauge theory, from instanton effects, and we have argued that vacuum energy is not consistent with an ultrasoft theory. But it is also known that the addition of massless fermions can have the effect of removing this instanton contribution to vacuum energy \cite{n6}.

Thus consider a confining QCD-like theory with massless quarks, which also experiences chiral symmetry breaking. The mass gap is again not infinitely soft due to the appearance of the quark condensate, which introduces power law corrections in various operator product expansions. But since the quark condensate is chiral, it must appear squared in operator products that do not carry chirality. In particular for the gluon propagator
\begin{equation}
\lim_{p^2\rightarrow\infty}[p^2 G(p^2)]^{-1}= f_G(p/\mu)(1+c\frac{\langle\overline{q}q\rangle^2}{p^6}+...)
.\end{equation}
This has the same behavior as the ultrasoft scalar field theory discussed above. That discussion then has a bearing on the self-consistency of this form for the gluon propagator. The suggestion is that the nonperturbative physics is producing a mass gap, and that it is doing so as softly as consistently possible.

A by-product of this picture is a vanishing vacuum energy and a vanishing gluon condensate. It leads to the question of whether the gluon condensate that is thought to exist in QCD would survive the limit of vanishing current quark masses. The experimental and lattice-based evidence for a gluon condensate in this limit was reviewed in \cite{n7} and the evidence was found to be lacking. It appears to be a question that deserves more attention, given the relation it has to our understanding of vacuum energy in theories of fundamental interest.

\section*{Acknowledgments}
I thank E.~Poppitz for his comments. This work was supported in part by the Natural Sciences and Engineering Research Council of Canada.


\begin{thebibliography}{11}
\bibitem{n1} L.~S.~Brown, Quantum Field Theory, Cambridge University Press, 1994.
\bibitem{n2} V.~N.~Gribov, Nucl.~Phys.~B139 (1978) 1; R.~F.~Sobreiro and S.~P.~Sorella, arXiv:hep-th/0504095. 
\bibitem{n3} M.A.~Semenov-Tyan-Shanskii and V.A.~Franke, Pubs. LOMI Seminar 120 (1982) 159, [J.~Sov.~Math.~34 (1986) 1999]. 
\bibitem{n4} B.~Holdom, Phys. Rev. D79 (2009) 085013, arXiv:0901.0497.
\bibitem{n5} D.~Zwanziger, Nucl.~Phys.~B323, 513 (1989).
\bibitem{n6} G. `t Hooft, Phys. Rev. D14 (1976) 3432.
\bibitem{n7} B.~Holdom, New J.~Phys.~10 No 5 (2008) 053040, arXiv:0708.1057.
\end{thebibliography}
\end{document}